\documentclass
[superscriptaddress,secnumarabic,amssymb,amsmath,nobibnotes,aps,prd,showkeys,nofootinbib,twocolumn,notitlepage,10pt]{revtex4}%
\usepackage{setspace}
\usepackage{color}
\usepackage{amsmath}
\usepackage{amsfonts}
\usepackage{verbatim}
\usepackage{amssymb}
\usepackage[colorlinks]{hyperref}
\usepackage{graphicx,bm}
\usepackage{amsmath}
\usepackage{amssymb}
\usepackage{graphicx}
\usepackage{textgreek}
\usepackage[caption=false]{subfig}%
\providecommand{\U}[1]{\protect\rule{.1in}{.1in}}
\newcommand{\be}{\begin{equation}}
\newcommand{\ee}{\end{equation}}

\newcommand{\mincir}{\raise
-3.truept\hbox{\rlap{\hbox{$\sim$}}\raise4.truept\hbox{$<$}\ }}
\newcommand{\magcir}{\raise
-3.truept\hbox{\rlap{\hbox{$\sim$}}\raise4.truept\hbox{$>$}\ }}

\ifx\pdfoutput\relax\let\pdfoutput=\undefined\fi
\newcount\msipdfoutput
\ifx\pdfoutput\undefined\else
\ifcase\pdfoutput\else
\ifx\paperwidth\undefined\else
\ifdim\paperheight=0pt\relax\else\pdfpageheight\paperheight\fi
\ifdim\paperwidth=0pt\relax\else\pdfpagewidth\paperwidth\fi
\fi\fi\fi
\begin{document}
\title{Challenging $\Lambda$CDM with Higher-Order GUP Corrections}
\author{Andronikos Paliathanasis}
\email{anpaliat@phys.uoa.gr}
\affiliation{Institute of Systems Science, Durban University of Technology, Durban 4000,
South Africa}
\affiliation{Centre for Space Research, North-West University, Potchefstroom 2520, South Africa}
\affiliation{National Institute for Theoretical and Computational Sciences (NITheCS), South Africa}
\affiliation{Departamento de Matem\`{a}ticas, Universidad Cat\`{o}lica del Norte, Avda.
Angamos 0610, Casilla 1280 Antofagasta, Chile}
\author{Genly Leon}
\email{genly.leon@ucn.cl}
\affiliation{Departamento de Matem\'{a}ticas, Universidad Cat\`{o}lica del Norte, Avda.
Angamos 0610, Casilla 1280 Antofagasta, Chile}
\affiliation{Institute of Systems Science, Durban University of Technology, Durban 4000,
South Africa}
\author{Yoelsy Leyva}
\email{yoelsy.leyva@academicos.uta.cl}
\affiliation{Departamento de F\'{\i}sica, Facultad de Ciencias, Universidad de Tarapaca,
Casilla 7-D, Arica, Chile}
\author{Giuseppe Gaetano Luciano}
\email{giuseppegaetano.luciano@udl.cat}
\affiliation{Departamento de Qu\'{\i}mica, F\'{\i}sica y Ciencias Ambientales y del Suelo,
Escuela Polit\'ecnica Superior -- Lleida, Universidad de Lleida, Av. Jaume II,
69, 25001 Lleida, Spain}
\author{Amare Abebe}
\email{amare.abebe@nithecs.ac.za}
\affiliation{Centre for Space Research, North-West University, Potchefstroom 2520, South Africa}
\affiliation{National Institute for Theoretical and Computational Sciences (NITheCS), South Africa}

\begin{abstract}
We study quantum corrections to the $\Lambda$CDM model arising from a minimum
measurable length in Heisenberg's uncertainty principle. We focus on a
higher-order Generalized Uncertainty Principle, beyond the quadratic form.
This generalized GUP introduces two free parameters, and we determine the
modified Friedmann equation. This framework leads to a perturbative
cosmological model that naturally reduces to $\Lambda$CDM in an appropriate
limiting case of the deformation parameters. We construct the modified
cosmological scenario, analyze its deviations from the standard case, and
examine it as a mechanism for the description of dynamical dark energy. To
constrain the model, we employ Cosmic Chronometers, the latest Baryon Acoustic
Oscillations from the DESI DR2 release, and Supernova data from the
PantheonPlus and Union3 catalogues. Our analysis indicates that the modified
GUP model is statistically competitive with the $\Lambda$CDM scenario,
providing comparable or even improved fits to some of the combined datasets.
Moreover, the data constrain the deformation parameter of the GUP model, with
the preferred value found to be negative, which corresponds to a phantom
regime in the effective dynamical dark energy description.

\end{abstract}
\date{\today}
\maketitle

\section{Introduction}

\label{sec1}

The recent cosmological data support cosmological theories that deviate from
the standard $\Lambda$-Cold Dark Matter ($\Lambda$CDM) model, with a dynamical
dark energy component that can cross the phantom divide
line~\cite{union,des4,des5,des6,ra1,ra2,ra3,ks1,ten1,your,orl}. There is a
plethora of proposed models in the literature that attempt to explain the
cosmological data from phenomenological and theoretical perspectives and
elucidate the origin of the dark energy component. Parametric dark energy
models were recently considered in~\cite{a1,a2,a3,a4,a5,a6,a66,a666,a6b,a6c},
non-cold dark matter models were investigated in~\cite{a77,a88,a99}, while
scalar fields and modified theories of gravity were examined
in~\cite{a7,a8,a9,a10,a10a,a10b,a11c,a11d}. Cosmological models with
interaction in the dark sector were studied
in~\cite{a11,a12,a13,a14,a15,a15a,supr,Pan:2025qwy}, and entropic theories
in~\cite{Sheykhi:2018dpn,Lymperis:2018iuz,Saridakis:2020lrg,Lymperis:2021qty,Nojiri:2022dkr,Gohar:2023hnb,a16,a17}%
. On the other hand, quantum effects in the dark sector of the universe were
examined in~\cite{a18,a19,a20,angup2}.

In this context, the Generalized Uncertainty Principle (GUP) emerges as a
quantum-gravity inspired modification of Heisenberg's uncertainty relation,
which incorporates a fundamental minimal length and gives rise to a deformed
algebraic structure~\cite{GUP1,GUP2,GUP3,GUP4}. Such a framework modifies the
dynamical laws by embedding quantum effects into the equations of motion. In
particular, as shown in~\cite{angup2}, the quadratic form of the GUP can be
phenomenologically recast within cosmology so as to yield semiclassical
deformations of the Friedmann equations in the $\Lambda$CDM scenario. The
resulting model admits a closed-form solution and provides an effective
description of dynamical dark energy capable of crossing the phantom divide line.

Over the past decades, numerous studies have investigated the implications of
a minimum length in gravitational
theories~\cite{de1,de2,de4,bh1,bh2,bh3,de4a,de4b,de4d,de5,de10,de9,de8,bh5,bh6,bb1,Jizba:2022icu,angup}%
. It has been shown that the impact of the GUP on cosmological field equations
exhibits dynamical similarities to those found in higher-order gravity
theories~\cite{f1}, as well as in scalar field models incorporating a
Pais-Uhlenbeck oscillator term~\cite{f2}. Additionally, various formulations
of the GUP have been proposed in the literature, each motivated by different
quantum-gravity considerations. These distinct versions provide alternative
phenomenological frameworks, enriching the exploration of minimum-length
effects in both gravitational and cosmological settings (see~\cite{BossoRev}
for a recent review).

Starting from the above premises, in this work we extend the analysis
presented in~\cite{angup2} by considering a more general GUP. Specifically, we
consider a two-parameter higher-order formulation and analyze its implications
for the modifications of the $\Lambda$CDM cosmology. We apply late-time
cosmological data to test the higher-order GUP as a potential mechanism for
dynamical dark energy behavior. This analysis allows us to assess the
implications of the GUP in cosmological studies and to identify the specific
form of its correction to Heisenberg's uncertainty relation. Finally, we use
cosmological data to constrain the GUP model.

The structure of the rest of the paper is as follows: in Sec.~\ref{sec2} we
introduce the GUP and study the effects of the minimum measurable length in
the classical field equations. The cosmological applications of the GUP are
discussed in Sec.~\ref{sec3}, where we review previous results for the
$\Lambda$CDM universe in a spatially flat
Friedmann-Lema\^{\i}tre-Robertson-Walker(FLRW) spacetime, in which the second
Friedmann equation is modified by the presence of the GUP. We then consider a
higher-order GUP model and reconstruct the master equation governing cosmic
evolution. This equation depends on two parameters: the deformation parameter
of the GUP and the exponent characterizing the higher-order extension. The
main results of this study are presented in Sec.~\ref{sec4}, where we
constrain our GUP-modified cosmological theory with late-time observational
data and compare it with the $\Lambda$CDM model. For the statistical analysis,
we consider Cosmic Chronometers, the recent data for the Baryon Acoustic
Oscillations from the DESI DR2 collaboration and the Supernova observations of
the PantheonPlus and Union3 catalogues. Finally, in Sec.~\ref{sec5} we draw
our conclusions. Unless otherwise specified, Planck units are used throughout.

\section{Generalized Uncertainty Principle}

\label{sec2}

The most widely employed realization of the GUP {in the literature is
the quadratic model considered in}~\cite{GUP3}, i.e.,
\begin{equation}
\Delta X\Delta P\geq\frac{1}{2}\left(  1+\beta_{0}\Delta P^{2}+\hat{\gamma
}\right)  ,\label{GUP}%
\end{equation}
where $\Delta X,\Delta P$ are the position and momentum uncertainties,
respectively, and $\beta_{0}$ is a suitable parameter with dimension $1/P^{2}$.

A generalization of Eq.~\eqref{GUP} can {be formulated by introducing
the following deformation of the canonical algebra \cite{mag1,mag2}}
\begin{equation}
\lbrack X_{i},P_{j}]=i\left[  \delta_{ij}(1-\beta_{0}P^{2})-2\beta_{0}%
P_{i}P_{j}\right]  ,\label{md1}%
\end{equation}
while the spatial coordinates commute, as do the momentum
components\footnote{The range of variability of the indices $i,j$ depends on
the specific GUP model considered. In~\cite{Moayedi}, for instance, a
Lorentz-covariant deformed algebra is proposed, with the indices
$i,j\in\{1,2,3,4\}$. A possible drawback of this model is that it also
involves a commutator for the time operator. However, the definition of such
operator in the quantum framework remains an open problem~\cite{Bauer:2016aio}%
. As our present goal is to investigate the phenomenological implications of
the minimal length in cosmology, we postpone a detailed analysis of this
conceptual aspect to future work.}. Deformations of this type have been
considered in the literature in the context of quantum gravity-motivated
models~\cite{Quesne2006,Vagenas,Kemph1,Kemph2,Vag1,Vag2,Moayedi}.
%Although in some of these approaches attempts have been made to construct a
%covariant formulation of the deformed commutator involving also the conjugate
%time–energy variables, the definition of a consistent quantum time operator
%remains an open problem~\cite{Bauer:2016aio}. In the present work, we shall not address this issue
%and restrict our attention to deformed commutators acting only on the spatial
%coordinates, with indices $i,j \in \{1,2,3\}$.
%By applying the Robertson–Schrödinger inequality, expression~(\ref{md1}) gives rise to the corresponding %generalized uncertainty relation~\cite{nnd1}.

In both Eqs.~\eqref{GUP} and~\eqref{md1} the parameter $\beta_{0}$ is the
deformation parameter, which is directly related to the existence of a minimum
measurable length. It is usually expressed as $\beta_{0}={\beta}/{M_{Pl}%
^{2}c^{2}=\beta}\ell_{Pl}^{2}/2\hbar^{2}$, where $M_{Pl}$ is the Planck mass,
$\ell_{Pl}$ ($\approx10^{-35}~m)$ is the Planck length, $M_{Pl}c^{2}$
($\approx1.2\times10^{19}~GeV)$ is the Planck energy and $\beta$ is a
dimensionless coupling parameter, often assumed to be of order unity in string
theory~\cite{GUP1,Amati:1987wq} and extended field theoretical models
\cite{hm5} (here we have temporarily restored the correct units). Although
$\beta$ is commonly taken to be positive, a number of studies have considered
the possibility of negative values of $\beta$ (see,
e.g.,~\cite{nn1,nn2,nn3,nn4,Jizba:2022icu}). Clearly, in the limit
$\beta\rightarrow0$ the standard Heisenberg algebra, together with the usual
structure of quantum mechanics, is recovered.

%The existence of the minimum length leads to the modified deformed algebra%
%\begin{align}
%\lbrack X_{i},P_{j}]  &  =i\hbar\lbrack\delta_{ij}(1-\beta P^{2})-2\beta
%P_{i}P_{j}],\label{md1}\\
%\left[  X_{i},X_{j}\right]   &  =0,\\
%\left[  P_{i},P_{j}\right]   &  =0.
%\end{align}

For computational purposes, we now adopt the coordinate representation for the
position and momentum operators. We set $X_{i} = x_{i}$ and modify the
momentum operator as~\cite{Moayedi}
\begin{equation}
P_{i}=(1-\beta_{0} p^{2})p_{i}, \label{sd1}%
\end{equation}
where $x_{i},~p_{i}$ denote the standard operators satisfying the canonical
algebra $[x_{i},p_{j}]=i\hbar\delta_{ij}$\thinspace$,$ and $p^{2}=p_{i}p^{i}$.

As a paradigmatic system to study the effects of the GUP, we consider a free
spin-0 particle $\Phi$ with rest mass $m$. For such a system, the Klein-Gordon
equation reads%
\begin{equation}
P_{i}P^{i}\Phi-m^{2}\Phi=0.
\end{equation}
Within the coordinate representation, making use of (\ref{sd1}) one
obtains~\cite{Moayedi}
\begin{equation}
\square\Phi+2\beta_{0}\square\left(  \square\Phi\right)  +m^{2}\Phi+O\left(
\beta_{0}^{2}\right)  =0,
\end{equation}
where $\square=\partial_{i}\partial^{i}$ denotes the d'Alembertian operator,
$\partial^{j}\equiv\frac{\partial}{\partial x_{j}}=-i p^{j}$ and $2\beta
_{0}\square\left(  \square\right)  $ is a fourth-order operator which
describes the quantum corrections to the GUP. Owing to the deformation of the
algebra, the emergence of a minimum length gives rise to an additional quantum
correction term in the classical field equations.

Let $\mathcal{H}=\mathcal{H}(x,p)$ be the Hamiltonian function describing the
classical system. In the classical limit, the equations of motion follow from
Hamilton's equations and are given by
\begin{equation}
\dot{x}_{i}=\{x_{i},\mathcal{H}\},\qquad\dot{p}_{i}=\{p_{i},\mathcal{H}\},
\end{equation}
where $\{\cdot,\cdot\}$ denotes the Poisson bracket. By expanding the bracket,
one obtains the modified Hamilton's equations~\cite{hm1,hm2,hm3}:
\begin{equation}
\dot{x}^{i}=\left(  1-\beta_{0}p^{2}\right)  \frac{\partial\mathcal{H}%
}{\partial p_{i}},\qquad\dot{p}^{i}=\left(  1-\beta_{0}p^{2}\right)
\frac{\partial\mathcal{H}}{\partial x_{i}}.
\end{equation}

Up to this point, we have considered the quadratic form of the GUP. However,
the notion of a minimum length can also arise within a more general framework,
in which the uncertainty relation takes the form~\cite{hm4,hm5,hm6,hm7,hm8}
\begin{equation}
\Delta X_{i}\Delta P_{j}\geqslant\frac{1}{2}[\delta_{ij}(1+\beta f\left(
P\right)  )],
\end{equation}
for suitable choices of the dimensionless function $f(P)$~\cite{BossoPW}.
%For dimensional consistency, the product $\beta f(P)$ must be dimensionless, which implies that $f(P)$ itself has to be dimensionless as well.

In this framework, Hamilton's equations governing the classical dynamics are
accordingly modified as
\begin{equation}
\dot{x}^{i}=\left(  1-\beta f(p)\right)  \frac{\partial\mathcal{H}}{\partial
p_{i}}~,~\dot{p}^{i}=\left(  1-\beta f(p)\right)  \frac{\partial\mathcal{H}%
}{\partial x_{i}}. \label{md3}%
\end{equation}

\section{$\Lambda$CDM with GUP Corrections}

\label{sec3}

In \cite{angup}, the effects of the GUP on the field equations of Szekeres
spacetimes were investigated. In particular, when the field equations for the
Szekeres system are expressed in terms of the propagation and constraint
equations~\cite{tsa1}, the propagation equations constitute a Hamiltonian
system. This allows quantum correction terms to be introduced into the
gravitational field equations.

The FLRW geometry is recovered in the limit where the Szekeres geometry
becomes isotropic and homogeneous. The propagation equations for the $\Lambda
$CDM model are
\begin{align}
\dot{\rho}_{m}+3H\rho_{m}  &  =0,\label{fe.01}\\
\dot{H}+\frac{H^{2}}{9}+3\left(  \frac{1}{2}\rho_{m}-\Lambda\right)   &  =0,
\label{fe.02}%
\end{align}
with the constraint
\begin{equation}
3H^{2} = \rho_{m} + \Lambda, \label{fe.03}%
\end{equation}
where $H=\dot a/a$ denotes the Hubble parameter, $a$ is the scale factor,
$\rho_{m}$ is the energy density of cold dark matter (CDM) and $\Lambda$
represents the cosmological constant (for our purposes of studying the GUP as
a potential mechanism for dynamical dark energy behavior, we can safely
neglect the radiation component). An overdot indicates differentiation with
respect to time.

Equations (\ref{fe.01}), (\ref{fe.02}) form a Hamiltonian system with
Hamiltonian function%
\begin{equation}
\mathcal{H}\left(  t,\rho_{m},P\right)  =\frac{1}{2}\rho_{m}^{\frac{8}{3}%
}P^{2}-\frac{3}{2}\rho_{m}^{\frac{1}{3}}\left(  1+\frac{\Lambda}{\rho_{m}%
}\right)  ,
\end{equation}
where the momentum is defined as $P=\dot{\rho}_{m}\rho_{m}^{-\frac{8}{3}}$.

In the presence of the minimum length, the modified Hamilton's equation lead
to the modified second Friedmann's equation%
\begin{equation}
\dot{H}+H^{2}=\frac{1}{6}\left(  2\Lambda-\rho_{m}\right)  \left(  1+2\beta
f\left(  t\right)  \right)  +\beta\dot{f}\left(  t\right)  H .
\end{equation}
Thus, the evolution of the fractional matter energy density,
\begin{equation}
\Omega_{m} = \frac{\rho_{m}}{3H^{2}}~, \label{fe.00}%
\end{equation}
is governed by the nonlinear differential equation
\begin{equation}
\frac{d\Omega_{m}}{d\ln a} = 3\left(  \Omega_{m}-1\right)  \Omega_{m} +
2\Omega_{m}\beta\!\left(  f(a)\left(  3\Omega_{m}-2\right)  - \frac
{df(a)}{d\ln a} \right)  , \label{fe.20}%
\end{equation}
or equivalently%
\begin{align}
-\left(  1+z \right)  \frac{d\Omega_{m}}{dz}  &  = 3\left(  \Omega_{m}-1
\right)  \Omega_{m} + 2\Omega_{m}\beta\Bigg( f(z)\left(  3\Omega_{m}-2 \right)
\nonumber\\
&  \quad+ \left(  1+z \right)  \frac{df(z)}{dz} \Bigg)\,, \label{fe.21}%
\end{align}
where $z = a^{-1}-1$ denotes the redshift (with $a_{0}=1$, where the subscript
$0$ refers to the present epoch).

Moreover, the equation-of-state parameter of the effective fluid is defined
as
\begin{equation}
w_{eff}=-1+\Omega_{m}+\frac{2}{3}\beta\left(  f\left(  z\right)  \left(
3\Omega_{m}-2\right)  +\left(  1+z\right)  \frac{df\left(  z\right)  }%
{dz}\right)  .
\end{equation}

From Eq.~(\ref{fe.21}), together with expression (\ref{fe.00}), we can
construct the Hubble function
\begin{equation}
\left(  \frac{H\left(  z\right)  }{H_{0}}\right)  ^{2}=\frac{\Omega
_{m0}\left(  1+z\right)  ^{3}}{\Omega_{m}\left(  z\right)  },\,\quad\Omega
_{m}\left(  0\right)  \equiv\Omega_{m0}\,, \label{fe.22}%
\end{equation}
where $\Omega_{m0}$ denotes the present-day matter energy density parameter.

Hence, the specification of the deformation function $f$ plays a central role
in governing the dynamics and evolution of the cosmological model.

\subsection{Quadratic GUP}

The model $f=\tfrac{H^{2}}{\rho_{m}}$ was introduced in~\cite{angup2}. The
factor $\tfrac{1}{\rho_{m}}$ was included to ensure that $f$ is correctly
dimensionless. Hence, when expressed in terms of the energy density parameter
$\Omega_{m}$, the function $f$ takes the form $f = \frac{1}{3\Omega_{m}}$.

For this GUP, the dynamics in Eq.~(\ref{fe.20}) becomes
\begin{equation}
\label{15}\frac{d\Omega_{m}}{d\ln a}=3\Omega_{m}\left(  \Omega_{m}-1\right)
+\left(  4\Omega_{m}-\frac{10}{3}\right)  \beta+O\left(  \beta^{2}\right)  ,
\end{equation}
which admits the following closed-form solution for the energy density:
\begin{equation}
\Omega_{m}\left(  a\right)  =\frac{1}{6}\left(  3-4\beta+\sqrt{9+16\beta
}-\frac{2\sqrt{9+16\beta\left(  1+\beta\right)  }}{1+\frac{1-\Omega
_{m0}a^{\sqrt{9+16\beta\left(  1+\beta\right)  }}}{\Omega_{m0}}}\right)  .
\end{equation}

The closed-form expression for the Hubble function is given as
\begin{equation}
H^{2}\left(  a\right)  =H_{\Lambda CDM}^{2}\left(  a\right)  +\beta
H_{cor}^{B}\left(  a\right)  +O\left(  \beta^{2}\right)  ,
\end{equation}
where $H_{cor}^{B}\left(  a\right)  $ encodes the quantum corrections to the
Hubble function, namely
\begin{align}
\frac{3H_{cor}^{B}\!\left(  a\right)  }{2H_{G0}^{2}} &  =\Omega_{m0}%
a^{-3}+6\left(  1-\Omega_{m0}\right)  \nonumber\\
&  \quad+\frac{5\left(  1-\Omega_{m0}\right)  ^{2}}{\Omega_{m0}}%
a^{3}+12\left(  1-\Omega_{m0}\right)  \ln a,\label{an.15}%
\end{align}
and $H_{G0}^{2}$ is a normalization constant. Building on this model, it was
shown in \cite{angup2} that GUP can serve as a mechanism for constructing
dynamical dark energy models. In particular, it has been employed as a
theoretical framework to realize a time-varying cosmological constant starting
from the standard one. It was further demonstrated that the resulting
GUP-modified $\Lambda$CDM model provides an improved fit to observational data
compared with the undeformed theory. 

{For the quadratic GUP, a closed-form solution for the Hubble function
was previously presented in \cite{vag01}. However, the solution obtained there
differs from the one considered in our work because we adopt a different
approach to formulating the GUP and modifying the gravitational field
equations. In \cite{vag01}, the authors modify Einstein's action integral by
replacing the standard momentum with its deformed part, which introduces
additional quantum-correction terms. In contrast, our analysis follows the
approach described in \cite{BossoPW}, where the deformed algebra is used to
modify Hamilton's equations, leading to new definitions for the observables,
including the Hubble function. These different formulations lead to different
reconstructed Hubble functions.}

\subsection{Beyond the quadratic GUP}

In the present study, we propose an extension of the above framework by
generalizing the functional dependence of the GUP model to $f=\left(
\frac{H^{2}}{\rho_{m}}\right)  ^{\gamma}$, where $\gamma$ is an additional
free parameter. The parameter $\gamma$ controls the strength and nonlinearity
of quantum corrections to the Friedmann equations, with $\gamma= 1$ reducing
to the quadratic GUP case.

This modification is motivated by both theoretical and phenomenological
considerations. On the theoretical side, the power-law form represents the
most natural generalization of the linear dependence previously assumed, with
the case $\gamma=1$ recovered as a special limit. It also allows for a broader
class of deformations that could emerge from different realizations of quantum
gravity, where non-linear corrections to the uncertainty principle may be
expected. On the phenomenological side, the inclusion of $\gamma$ provides
enhanced flexibility in modeling dynamical dark energy, as the ratio
$H^{2}/\rho_{m}$ encodes the interplay between the cosmic expansion rate and
the matter sector. By varying $\gamma$, one can interpolate between different
scaling behaviors of the effective cosmological term, thereby capturing a
richer range of cosmological evolutions.

In this framework, the evolution of the energy density is given by the
differential equation
\begin{align}
-\left(  1+z\right)  \frac{d\Omega_{m}}{dz} &  =3\Omega_{m}\left(  \Omega
_{m}-1\right)  \nonumber\label{fe.23}\\
&  \quad+2\left(  3\Omega_{m}\right)  ^{1-\gamma}\left(  (1+\gamma)\Omega
_{m}-\tfrac{2}{3}-\gamma\right)  \beta\,,
\end{align}
It is worth noting that for $\gamma=1$, Eq.~\eqref{fe.23} consistently
reproduces the dynamics of Eq.~\eqref{15}. Moreover, in the absence of a
minimal length, i.e.\ for $\beta=0$, the standard $\Lambda$CDM limit is recovered.

For arbitrary values of the parameter $\gamma$, however, Eq.~\eqref{fe.23}
does not admit a closed-form analytic solution. In this case, we solve the
differential equation numerically with the initial condition $\Omega
_{m}(0)=\Omega_{m0}$, and, using the definition in Eq.~\eqref{fe.22}, we
subsequently compute the Hubble function $H(z)$.

\section{Observational Constraints}

\label{sec4}

We investigate the higher-order GUP cosmology~(\ref{fe.23}) as a possible
mechanism for describing late-time cosmic acceleration. In particular, we
employ observational data to constrain the free parameters $\{H_{0},
\Omega_{m0}, \beta, \gamma\}$ of the model~(\ref{fe.23}), and we compare its
performance with that of the standard $\Lambda$CDM scenario.

\subsection{Observational Data}

In the following, we present the observational datasets employed in this study.

\begin{itemize}
\item Observational Hubble Data (OHD): We consider the Cosmic Chronometers
(CC), which provide direct measurements of the Hubble parameter without
relying on any cosmological assumptions. Cosmic Chronometers correspond to
passively evolving galaxies with synchronous stellar populations and similar
cosmic evolution~\cite{co01}. This dataset is therefore model independent. In
the present analysis, we use the 31 direct measurements of the Hubble
parameter in the redshift range $0.09 \leq z \leq1.965$ reported in~\cite{cc1}.

\item Baryonic acoustic oscillations (BAO): We employ the recent release of
the Dark Energy Spectroscopic Instrument (DESI DR2) baryon acoustic
oscillation (BAO) observations~\cite{des4,des5,des6}. This dataset provides
measurements of the transverse comoving angular distance ratio,
\begin{equation}
\frac{D_{M}}{r_{d}} = \frac{D_{L}}{\left(  1+z\right)  r_{d}},
\end{equation}
the volume-averaged distance ratio,
\begin{equation}
\frac{D_{V}}{r_{d}} = \frac{\left(  z D_{H} D_{M}^{2}\right)  ^{1/3}}{r_{d}}.
\end{equation}
and the Hubble distance ratio,
\begin{equation}
\frac{D_{H}}{r_{d}} = \frac{c}{r_{d} H(z)},
\end{equation}
at seven distinct redshifts, where $D_{L}$ refers to the luminosity distance
and $r_{d}$ denotes the sound horizon at the drag epoch, which in our analysis
is treated as a free parameter (we have restored $c$ for consistency with the
notation in the literature).

\item Supernova of PantheonPlus (PP): This catalogue includes 1701 light
curves of 1550 spectroscopically confirmed supernova events. The data provide
measurements of the distance modulus $\mu^{{obs}}$ at redshifts in the range
$10^{-3} < z < 2.27$~\cite{pan}. The theoretical distance modulus is defined
as
\begin{equation}
\mu^{{th}} = 5 \log_{10} D_{L} + 25,
\end{equation}
where, in a spatially flat FLRW geometry, the luminosity distance is expressed
in terms of the Hubble function as $D_{L}(z) = c \,(1+z) \int_{0}^{z}
\frac{dz^{\prime}}{H(z^{\prime})}$. We employ the PantheonPlus catalogue
without the SH0ES Cepheid calibration.

\item Supernova of Union3 (U3): This Supernova catalogue includes 2087 events
within the same redshift range as the PP data, of which 1363 are shared with
the PantheonPlus catalogue \cite{union}.
\end{itemize}

\subsection{Methodology}

For the analysis of our cosmological theory, we employ the Bayesian inference
framework COBAYA\footnote{https://cobaya.readthedocs.io/}~\cite{cob1,cob2},
using a custom theory implementation together with the MCMC
sampler~\cite{mcmc1,mcmc2}. Furthermore, for the analysis of the results, we
make use of the GetDist library\footnote{https://getdist.readthedocs.io/}%
~\cite{getd}.

We also apply the same observational tests to the $\Lambda$CDM model. Given
the different number of free parameters, we use the Akaike Information
Criterion (AIC)~\cite{AIC} to assess which model is statistically favored. If
$\chi_{\min}^{2}=-2\ln\mathcal{L}_{\max}$ denotes the minimum chi-squared
value corresponding to the maximum likelihood, then the AIC is defined as
\begin{equation}
{AIC} \simeq-2\ln\mathcal{L}_{\max} + 2\kappa,
\end{equation}
where $\kappa$ represents the number of free parameters of the model.

For the higher-order GUP cosmology, $\kappa=5$, and for the $\Lambda$CDM model
is $\kappa=3$. Consequently,
\begin{equation}
\Delta AIC=AIC_{GUP}-AIC_{\Lambda}%
\end{equation}
which can be equivalently expressed as
\begin{equation}
\Delta{AIC} = \chi^{2}_{{GUP},\,\min} - \chi^{2}_{\Lambda,\,\min} + 4.
\end{equation}

According to Akaike's scale, the value of $\Delta AIC$ provides information on
which model offers a better fit to the data. Specifically, for $\lvert
\Delta{AIC} \rvert< 2$, the two models are statistically equivalent; for $2 <
\lvert\Delta{AIC} \rvert< 6$, there is weak evidence in favor of the model
with the smaller AIC value; if $6 < \lvert\Delta{AIC} \rvert< 10$, the
evidence is strong; and for $\lvert\Delta{AIC} \rvert> 10$, there is clear
evidence supporting the preference for the model with the lower AIC.

At this stage, it is important to note that, in order to eliminate the
influence of systematic errors in the comparison between the two models, the
Hubble function for the $\Lambda$CDM model has also been derived numerically,
using the same procedure as for Eq.~(\ref{fe.23}).%

%TCIMACRO{\TeXButton{B}{\begin{table}[tbp] \centering}}%
%BeginExpansion
\begin{table}[tbp] \centering
%EndExpansion
\caption{Priors for the Free Parameters}%
\begin{tabular}
[c]{ccc}\hline\hline
\textbf{Priors of Free Parameters} & $\mathbf{\,GUP}$ & $\mathbf{\Lambda}%
$\textbf{CDM}\\\hline
$\mathbf{H}_{0}~\left[  km~s^{-1}~Mpc^{-1}\right]  $ & $\left[  60,80\right]
$ & $\left[  60,80\right]  $\\
$\mathbf{\Omega}_{m0}$ & $\left[  0.01,0.99\right]  $ & $\left[
0.01,0.99\right]  $\\
$\mathbf{r}_{d}~\left[  Mpc\right]  $ & $\left[  120,170\right]  $ & $\left[
120,170\right]  $\\
$\mathbf{\beta}$ & $\left[  -0.3,0.3\right]  $ & $\nexists$\\
\textbf{$\gamma$} & $\left[  -10,10\right]  $ & $\nexists$\\\hline\hline
\end{tabular}
\label{tab2}%
%TCIMACRO{\TeXButton{E}{\end{table}}}%
%BeginExpansion
\end{table}%
%EndExpansion

\subsection{Results}%

%TCIMACRO{\TeXButton{B}{\begin{table}[tbp] \centering}}%
%BeginExpansion
\begin{table*}[tbp] \centering
%EndExpansion
\caption{Best-Fit Parameters for the Higher-Order GUP theory}%
\begin{tabular}
[c]{ccccc}\hline\hline
\textbf{Best Fit/Data} & \textbf{PP\&OHD} & \textbf{PP\&OHD\&BAO} &
\textbf{U3\&OHD} & \textbf{U3\&OHD\&BAO}\\\hline
$\mathbf{H}_{0}$ & $67.6_{-1.8}^{+1.8}$ & $68.1_{-1.7}^{+1.7}$ &
$66.3_{-1.9}^{+1.9}$ & $66.4_{-1.9}^{+1.9}$\\
$\mathbf{\Omega}_{m0}$ & $0.319_{-0.030}^{+0.047}$ & $0.307_{-0.014}^{+0.019}$
& $0.333_{-0.035}^{+0.051}$ & $0.322_{-0.017}^{+0.015}$\\
$\mathbf{r}_{d}$ & $\nexists$ & $147.1_{-3.4}^{+3.4}$ & $\nexists$ &
$147.3_{-3.8}^{+3.2}$\\
$\mathbf{\beta}$ & $0.002_{-0.1}^{+0.1}$ & $-0.031_{-0.030}^{+0.038}$ &
$-0.034_{-0.035}^{+0.046}$ & $-0.0564_{-0.0025}^{+0.045}$\\
\textbf{$\gamma$} & $1.3_{-3.5}^{+1.3}$ & $1.8_{-3.2}^{+1.1}$ & $>0.254$ &
$3.7_{-4.2}^{+2.0}$\\
$\mathbf{\chi}_{GUP\min}^{2}\mathbf{-\chi}_{\Lambda\min}^{2}$ & $-1.0$ &
$-3.33$ & $-3.19$ & $-7.61$\\
$\mathbf{AIC}_{GUP}\mathbf{-AIC}_{\Lambda}$ & \thinspace$+3.0$ & $+0.67$ &
$+0.81$ & $-3.61$\\\hline\hline
\end{tabular}
\label{tab3}%
%TCIMACRO{\TeXButton{E}{\end{table}}}%
%BeginExpansion
\end{table*}%
%EndExpansion

For the MCMC sampling, we adopt the priors listed in Table~\ref{tab2}. The
results obtained for each dataset are presented below.

For the combined PP\&OHD dataset, we obtain the best-fit parameters
$H_{0}=67.6_{-1.8}^{+1.8}\hspace{1mm}km\hspace{0.6mm}s^{-1}Mpc^{-1}$,
$\Omega_{m0}=0.319_{-0.030}^{+0.047}$, $\beta=0.002_{-0.1}^{+0.1}$, and
$\gamma=1.3_{-3.5}^{+1.3}$. The comparison with the $\Lambda$CDM model yields
$\chi_{{GUP},\,\min}^{2}-\chi_{\Lambda,\,\min}^{2}=-1.0$ and $\Delta
{AIC}=+3.0$, which indicates weak evidence in favor of the $\Lambda$CDM scenario.

When the BAO data are included, i.e., for the combined dataset PP\&OHD\&BAO,
the best-fit parameters are $H_{0}=68.1_{-1.7}^{+1.7}\hspace{1mm}%
km\hspace{0.6mm}s^{-1}Mpc^{-1}$, $\Omega_{m0}=0.307_{-0.014}^{+0.019}$,
$\beta=-0.031_{-0.030}^{+0.038}$ and $\gamma=1.8_{-3.2}^{+1.1}$. In this case,
$\chi_{GUP\min}^{2}-\chi_{\Lambda\min}^{2}=-3.33$ and $\Delta AIC=+0.67$.
Thus, the GUP model yields a slightly better fit to the data than $\Lambda
$CDM. However, once the larger number of free parameters is accounted for, the
AIC indicates that the two model are statistical equivalent.

For the U3\&OHD dataset, the best-fit parameters are $H_{0}=66.3_{-1.9}%
^{+1.9}\hspace{1mm}km\hspace{0.6mm}s^{-1}Mpc^{-1}$, $\Omega_{m0}%
=0.333_{-0.035}^{+0.051}$, $\beta=-0.034_{-0.035}^{+0.046}$ and $\gamma
>0.254$, with $\chi_{GUP\min}^{2}-\chi_{\Lambda\min}^{2}=-3.19$ and $\Delta
AIC=+0.81$. GUP model provides a slightly lower $\chi^{2}$ than $\Lambda$CDM,
the AIC once again indicates that there is not any preferred model by the data.

Finally, for the combined U3\&OHD\&BAO data, the best-fit parameters are
$H_{0}=66.4_{-1.9}^{+1.9}\hspace{1mm}km\hspace{0.6mm}s^{-1}Mpc^{-1}$,
$\Omega_{m0}=0.322_{-0.017}^{+0.015}$, $\beta=-0.0564_{-0.0025}^{+0.045}$, and
$\gamma=3.7_{-4.2}^{+2.0}$. The corresponding statistical indicators are
$\chi_{GUP\min}^{2}-\chi_{\Lambda\min}^{2}=-7.61$ and $\Delta AIC=-3.61$. In
this case, the GUP model provides a noticeably better fit to the data, and the
AIC indicates that GUP model has a week support over the $\Lambda$CDM from
this dataset.

Regarding the deformation parameter $\beta$, in most combined datasets the
value $\beta=0$ lies within $1\sigma$, except for U3\&OHD\&BAO, where it is
consistent only within $2\sigma$. For PP\&OHD, the best-fit value is $\beta
>0$, while for the other datasets it is $\beta<0$. According to
Ref.~\cite{angup2}, a negative $\beta$ is associated with phantom-like
behavior in the effective dark energy sector, leading to a more rapid cosmic
expansion. This interpretation is consistent with the U3\&OHD\&BAO results,
which strongly favor an accelerated expansion.

As for the parameter $\gamma$, the best-fit values are constrained to be
positive, with the case $\gamma=1$ lying within $1\sigma$. Nevertheless, for
most datasets the best-fit value is greater than unity, which points towards a
preference for the generalized GUP framework introduced in this work.

\begin{figure}[ptbh]
\centering\includegraphics[width=0.48\textwidth]{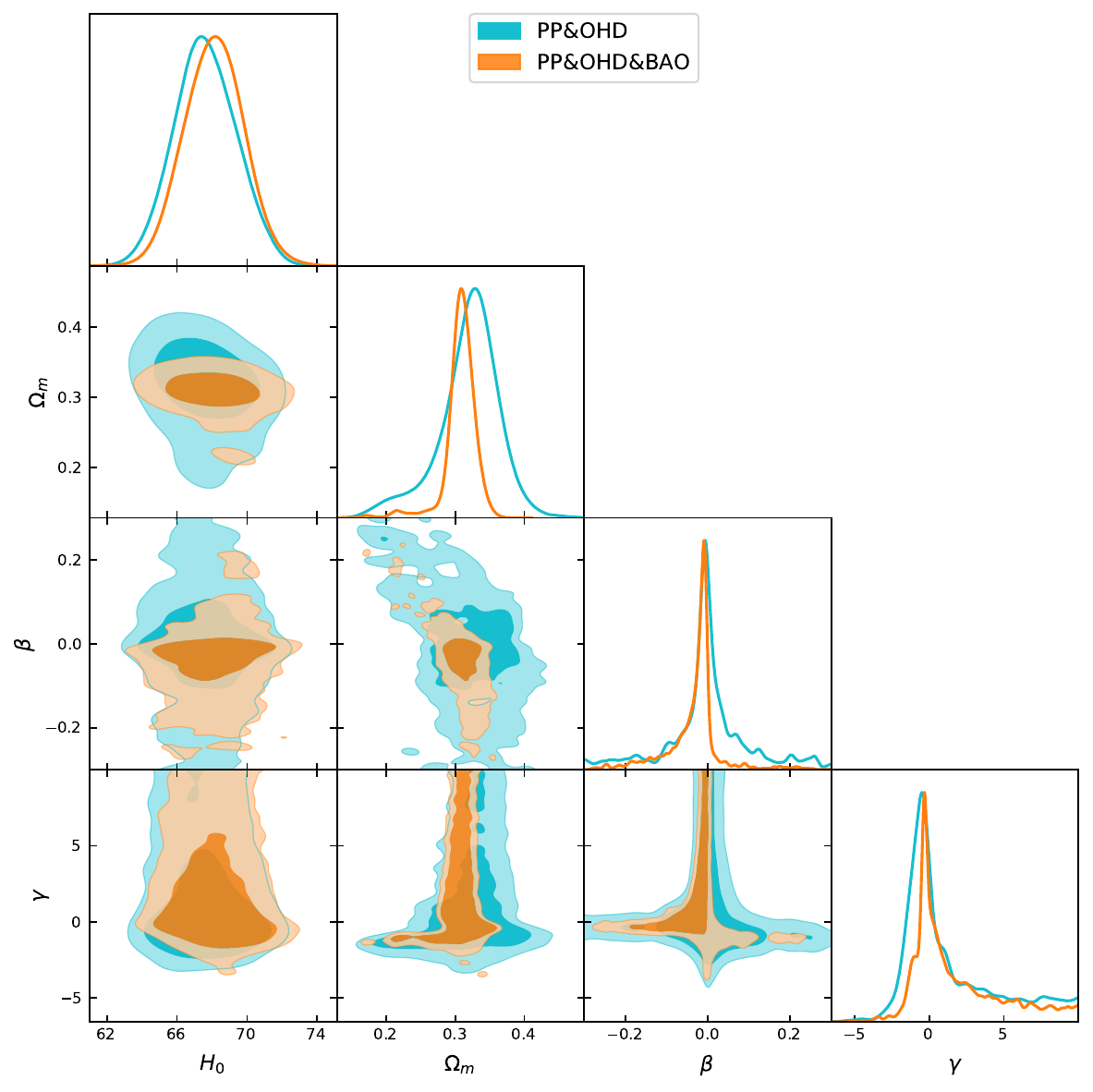}\caption{Confidence
regions of the posterior parameters for the GUP-modified model using the
datasets PP\&OHD and PP\&OHD\&BAO.}%
\label{fig1}%
\end{figure}

\begin{figure}[ptbh]
\centering\includegraphics[width=0.48\textwidth]{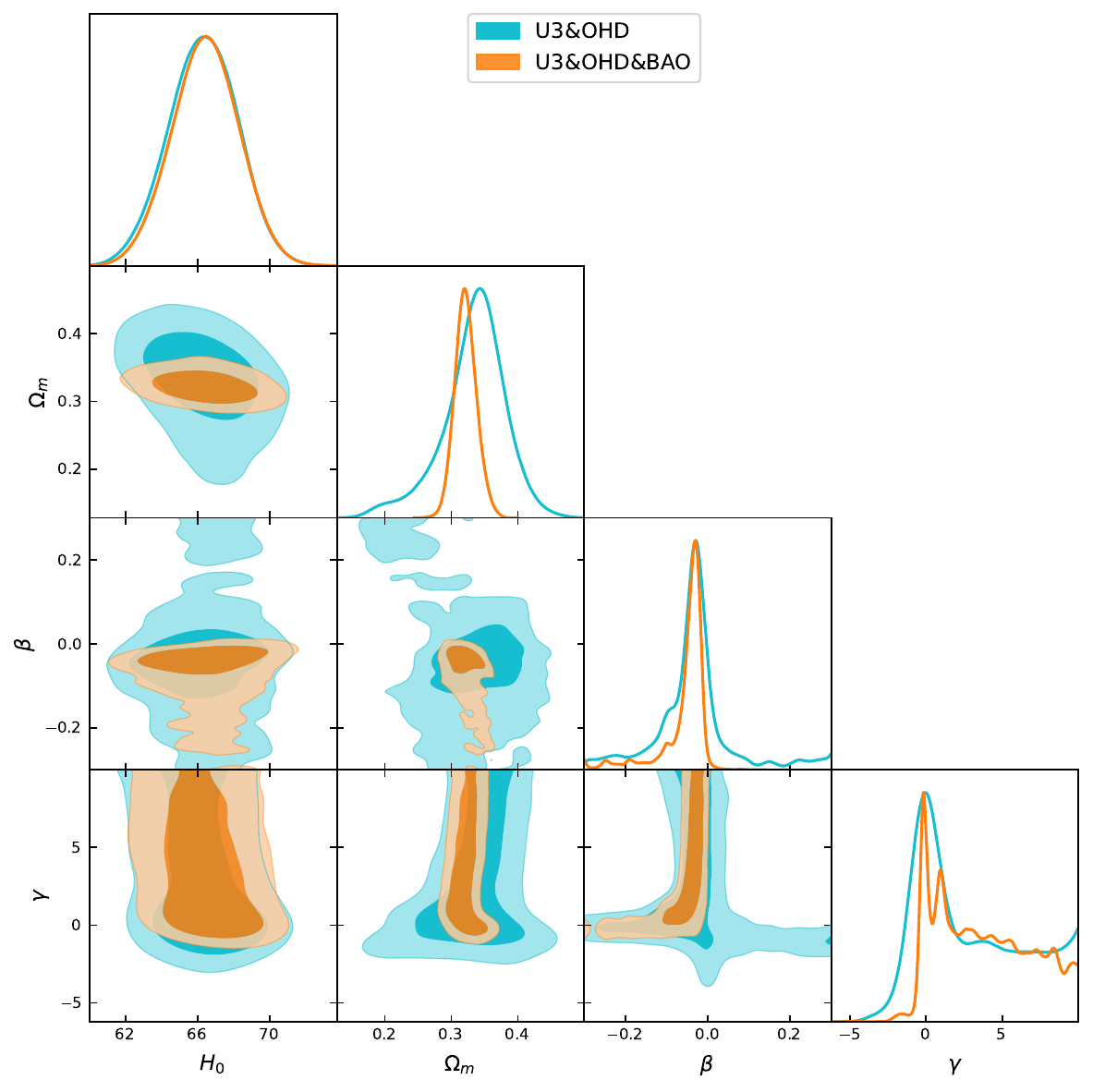}\caption{Confidence
space for the posterior parameters for the GUP-modified model with for the
datasets U3\&OHD and U3\&OHD\&BAO}%
\label{fig2}%
\end{figure}

In Figs.~\ref{fig1} and \ref{fig2}, we present the $2\sigma$ confidence
regions for the posterior parameters. In all cases, the best-fit values lie
within the $1\sigma$ contours. A summary of these results is provided in
Table~\ref{tab3}.

\section{Conclusions}

\label{sec5}

We examined the GUP as a possible mechanism to explain the late-time
acceleration of the universe. Focusing on a spatially flat FLRW geometry with
cold dark matter and a cosmological constant, we examined the modifications
induced by the existence of a minimal length scale in the classical
gravitational equations. We found that the second Friedmann equation - namely,
the Raychaudhuri equation - is altered in this framework, leading to a
modified $\Lambda$CDM cosmology that can naturally realize a dynamical dark
energy behavior.

For the GUP relation, we considered a two-parameter extension beyond the
quadratic (i.e., $\gamma=1$) case, incorporating higher-order derivative
corrections. The resulting cosmological field equations were solved
numerically and confronted with late-time observations, namely the
Observational Hubble Data from Cosmic Chronometers, the Baryon Acoustic
Oscillations from the DESI DR2 release and the Supernova samples from the
PantheonPlus and Union3 catalogues.

We fitted our modified cosmological model to different combinations of
observational data and found that the GUP-modified theory can reproduce the
observations with a quality comparable to, and in some cases slightly better
than, the standard $\Lambda$CDM scenario. Nevertheless, when model selection
criteria are applied, $\Lambda$CDM remains favored: owing to its smaller
number of free parameters, it achieves a lower AIC value. Consequently, the
Akaike Information Criterion does not permit us to draw firm conclusions
regarding a preferred model.

For all data combinations, the quadratic GUP limit is recovered within the
$1\sigma$ range of the parameter $\gamma$. However, the best-fit values of
$\gamma$ are greater than unity, lending support to the higher-order GUP
framework. In particular, when the Union3 data are included, the best-fit
value of $\gamma$ is further increased. Regarding the deformation parameter
$\beta$, the value $\beta=0$, which corresponds to the standard $\Lambda$CDM
model, lies within $1\sigma$ for the combined datasets PP\&OHD, PP\&OHD\&BAO,
and U3\&OHD, and within $2\sigma$ for the U3\&OHD\&BAO dataset. In addition,
our analysis reveals a tendency toward negative values, $\beta<0$. As
discussed in Ref.~\cite{angup2}, a negative deformation parameter $\beta$
leads the effective dynamical dark energy sector to exhibit phantom-like behavior.

In summary, the GUP offers a simple and natural mechanism through which
quantum effects can induce dynamical behavior in the dark energy sector.
Moreover, we have shown that cosmological observations of the expansion
history can be employed to constrain the free parameters of the deformation
algebra and, in particular, to place bounds on the deformation parameter
$\beta$.

In a future work, we plan to extend the present analysis by incorporating
Cosmic Microwave Background (CMB) data. While in this study we have focused on
low- and intermediate-redshift probes, the inclusion of CMB measurements will
provide tight constraints on the model parameters across the entire expansion
history. This will allow us to examine more rigorously whether the
higher-order GUP framework can alleviate existing cosmological tensions, such
as the discrepancy between local and early-universe determinations of the
Hubble constant, as well as possible deviations in the growth of cosmic structures.

\begin{acknowledgments}
AP \& GL authors thanks the support of VRIDT through Resoluci\'{o}n VRIDT No.
096/2022 and Resoluci\'{o}n VRIDT No. 098/2022. Part of this study was
supported by FONDECYT 1240514. The research of GGL is supported by the
postdoctoral fellowship program of the University of Lleida. GGL gratefully
acknowledges the contribution of the LISA Cosmology Working Group (CosWG), as
well as support from the COST Actions CA21136 - \textit{Addressing
observational tensions in cosmology with systematics and fundamental physics
(CosmoVerse)} - CA23130, \textit{Bridging high and low energies in search of
quantum gravity (BridgeQG)} and CA21106 - \textit{COSMIC WISPers in the Dark
Universe: Theory, astrophysics and experiments (CosmicWISPers)}.
\end{acknowledgments}

\end{document}